%% file: gioia.tex
\documentclass[graybox]{svmult}

\usepackage{mathptmx}       
\usepackage{helvet}         
\usepackage{courier}        
\usepackage{type1cm}        
\usepackage{makeidx}         
\usepackage{graphicx}        
\usepackage{multicol}        
\usepackage[bottom]{footmisc}

\makeindex             

\def\sun{\hbox{$\odot$~}}

\begin{document}

\title*{X-ray Clusters of Galaxies as Cosmological Tools}
\author{Isabella Maria Gioia}
\institute{Isabella Maria Gioia \at INAF-Istituto di Radioastronomia, 
Via Gobetti 101, 40129, Bologna, Italy, \\
\email{gioia@ira.inaf.it}
}

\maketitle

This paper was written as part of a book  entitled: ``Questions of 
Modern Cosmology - Galileo's Legacy'' which is a celebrative book 
dedicated to Galileo Galilei. The book is published in 2009, the
International Year of Astronomy, since it is intended to be a modern
tribute to the astronomer who, 400 years ago, first pointed a
telescope towards the night sky. The book is written in the form of
interviews between the editors and many physicists, astrophysicists
and cosmologists from all over the world. The editors engaged in
several discussions on the formation and evolution of the Universe
with the aim of summarizing the most important and significative
advances made by cosmology over the past century and at the beginning
of the new millennium. This paper deals with X-ray clusters of galaxies 
and how they can be used to constrain fundamental cosmological parameters.

\runinhead{Question:} Clusters of galaxies are powerful X-ray emitters
that can be easily detected out to high redshifts, and therefore are very
important tools for cosmologists. In particular, the relation between
the X-ray luminosity and the temperature, and temperature and mass of
the Intra Cluster Medium (ICM) offers a way to convert a luminosity
function into a mass function with obvious consequences for
present day cosmology. Can you comment on the state-of-the-art of
these studies? How can they constrain cosmological parameters?

\medskip\noindent
The properties of clusters are investigated through a multi-frequency
approach: from radio emission and Sunyaev-Zel'dovich effect, to
IR-Opt-UV mapping, to X-ray emission. For example, a cluster typically
appears more extended when mapped through the Sunyaev-Zel'dovich effect
than through its X-ray emission. Can you discuss the wealth of
astrophysical information achievable through the comparison among data
at various bands?  What advantages for cosmology come from such a
multi-wavelength approach?

\runinhead{Answer:}
To answer my cosmology question I need to give a little introduction
on why clusters of galaxies have always been a preferred tool of 
cosmologists. I will start with a short description of what clusters are, 
give a quick background on the different wavelengths, and describe
why X-ray clusters play such an important role in astronomy. Finally, 
I will give some description of the data that can be collected
at the different wavelengths and  discuss the advantages of the 
multi-wavelength approach for cosmology, and what needs to be done 
to improve our understanding of galaxy clusters.

\section{Clusters of Galaxies: an Introduction}
\label{sec:1}
\begin{figure}[b]
\sidecaption[b]
\includegraphics[scale=.43]{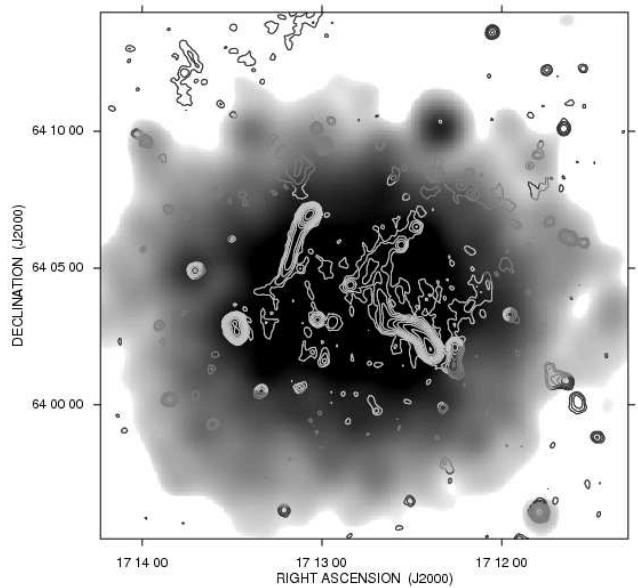}
\caption{This composite image contains two views of the cluster
Abell 2255. Superimposed onto the {\em ROSAT}-PSPC X-ray emission \cite{fer97}
(in shades of grey) are the VLA 1.4 GHz radio emission \cite{gov05}
represented as iso-contours. Courtesy of F. Govoni and M. Murgia.}
\label{fig:1}    
\end{figure}
\begin{figure}[t]
\sidecaption[t]
\includegraphics[scale=.43]{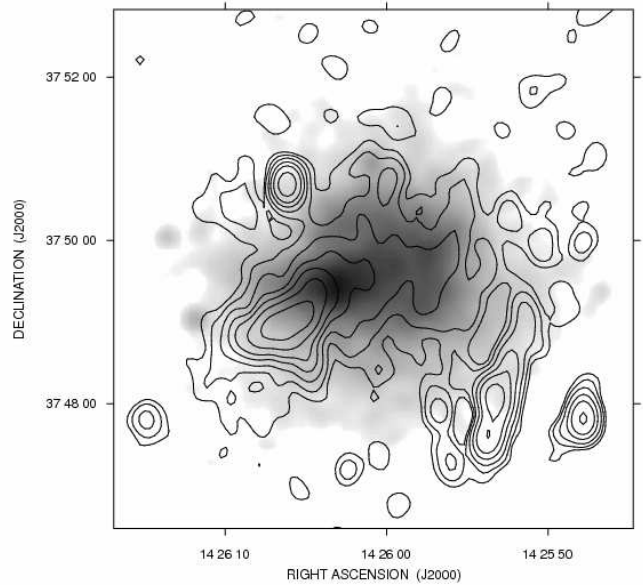}
\caption{This composite image contains two views of the cluster
Abell 1914. Superimposed onto the {\em Chandra} X-ray emission \cite{gov04}
(in shades of grey) are the VLA 1.4 GHz radio emission \cite{bach03}
represented as iso-contours. Courtesy of F. Govoni and M. Murgia.}
\label{fig:2}   
\end{figure}
\noindent 
About thirty years ago all clusters of galaxies were selected at
optical wavelengths since the easiest way to identify a cluster
is to search for an overdensity in the projected distribution of
galaxies in optical images. The pioneering work of
Abell \cite{a58} and later the catalogs by Zwicky and collaborators
\cite{zh61} made astronomers aware of how
many concentrations of galaxies were present in the nearby Universe
(that I define here as objects with a redshift z $\leq$ 0.15).
However, the most visible part of galaxy clusters, all of the stars in
all of the galaxies that make up the cluster, contributes only a small 
fraction of the mass of the cluster.  Clusters host manifold
components, such as individual galaxies and hot gas (the baryonic
component), invisible dark matter, and what are commonly referred to
as ``non thermal components''.  It is well known that a fraction of
clusters (about 40\% among rich, hot clusters \cite{gg99}) shows large
scale synchrotron radio emission with no obvious connection to the
cluster galaxies, and  therefore associated with the ICM (see
\cite{fer05} for a review on the subject). Such extended radio sources
are a direct and clear probe of the existence of cluster scale
magnetic fields and relativistic particles spread over the same large
volume. The composite images in Fig.~\ref{fig:1} and 
Fig~\ref{fig:2} are an illustration of the different features
seen in radio with the Very Large Array and in X-rays  with 
the {\em ROSAT}-PSPC and with {\em Chandra}, for two clusters, 
Abell 2255 and  Abell 1914.

\medskip\noindent
The total mass of a cluster is dominated by the non-baryonic component
(the invisible dark matter) that we know exists because of its
gravitational pull on the luminous matter. While the baryonic
component of a cluster can be directly observed at optical and X-ray
wavelengths, the invisible dark matter can only be measured through
the effect of gravitational lensing on the background galaxies or
observing other dynamical manifestations of the clusters.  Roughly, it
is estimated that the composition of a cluster is 3\% galaxies, 17\%
Intra Cluster Medium (ICM), and 80\% dark matter. Thus the total mass
of a cluster, which is the property we need to know to use clusters
as cosmological tools, is dominated by the invisible, collisionless
dark matter.

\section{Clusters of Galaxies in X-rays}
\label{sec:2}
Observations of galaxy clusters in the X-ray band trace the
intracluster gas, and hence provide an efficient and physically
meaningful method for the identification and selection of galaxy
clusters.  Over the past decade, studies based on the current
generation of X-ray satellites ({\em Chandra} and XMM-{\em Newton})
have completely changed our X-ray view of galaxy clusters. The large
collecting area of XMM-{\em Newton}, combined with the very fine angular
resolution of {\em Chandra}, have contributed to unveiling the complex
structure and physics of the hot ICM.

\medskip\noindent
The physics of X-ray emission from clusters of galaxies is pretty
straightforward.  Simple gravitational processes dominate cluster
formation and evolution and imply that clusters are still forming today.  
The evolution of clusters is simple, being driven by the gravity of the
underlying mass density field of the Universe and of a collisionless
collapse of the cluster dark matter.  These same formation processes also
heat gas trapped by the cluster potential, which then produces
optically thin thermal radiation. The evolution of cluster X-ray
emission can be more reliably calculated compared to that of other
objects visible at cosmological distances, such as galaxies and
quasars, and the cluster evolution calculations may be verified by
direct observations of nearby objects.  Thus observations of the X-ray
evolution of clusters provide a robust measure of the evolution of
cosmic structure and therfore constrain the cosmology of the Universe.

\medskip\noindent The advent of X-ray imaging in the 80's revealed
that clusters are extended and powerful sources, with luminosities up
to 10$^{45}$ erg s$^{-1}$ for the richest clusters, that emit by
optically thin thermal bremsstrahlung from hot ($\sim10^8$K),
low-density ($\sim10^{-3}$ atoms cm$^{-3}$) gas.  Their total masses
are in the range from a few times 10$^{13}$ M\sun for the poorest
groups to more than 10$^{15}$ M\sun for the most massive clusters. In
the X-ray sky clusters appear as high contrast objects, given the
dependence of the X-ray emission on the square of the gas density, and
can be seen up to high redshift.  In addition the X-ray luminosity,
L$_{X}$, correlates well with the cluster mass, the cluster property
most directly related to cosmological parameters (even though, as I
will discuss later, L$_{X}$ is not the most accurate of all proposed
X-ray indicators for the total mass of a cluster).

\begin{figure}[b]
\sidecaption[b]
\includegraphics[scale=.44]{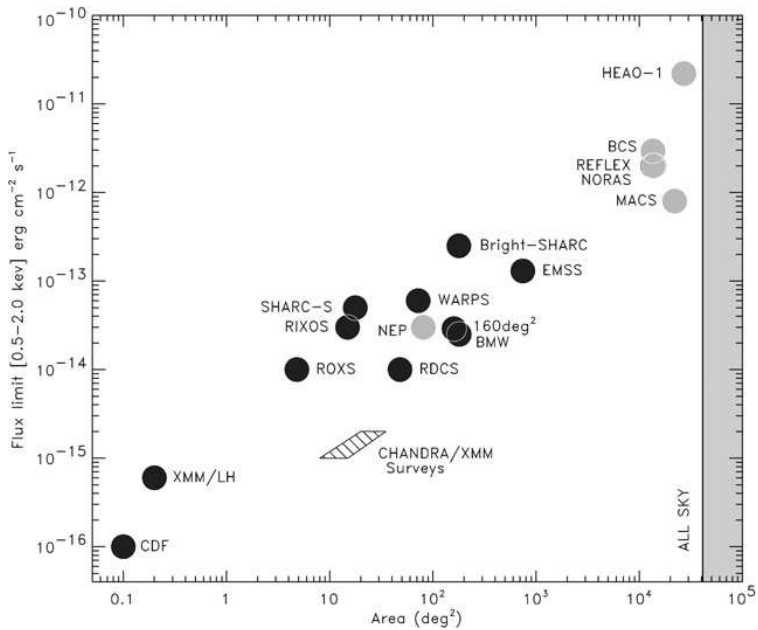}
\caption{Solid angles and flux-limits of X-ray surveys carried out
over the last two decades. Dark filled circles represent serendipitous
surveys constructed from a collection of pointed observations.  Light
shaded circles represent surveys covering contiguous area. The hatched
region is a predicted locus of serendipitous surveys with
\textit{Chandra} and XMM-\textit{Newton}. Figure comes from Rosati,
Borgani and Norman 2002.}
\label{fig:3}    
\end{figure}

\medskip\noindent Since the early 90's searches for clusters in X-ray
surveys discovered many bound systems out to cosmologically
interesting distances (see the pioneering work by Gioia and
collaborators with the {\em Einstein Observatory}
\cite{gio90a,gio90b,hen92}, and the many X-ray surveys that came out
later with the {\em ROSAT}-PSPC detector (cf. among others
\cite{ros95,vik98,gio03,bur07}). X-ray selection has the unique
advantage of revealing physical objects, deep potential wells in the
case of clusters, thus avoiding the problem of contamination by
foreground galaxies and stars as can happen with optical
selection. This is a fundamental point, especially when one deals with
very distant clusters which are the main players in cosmological
studies. An additional fundamental advantage of X-ray selection is the
ability to define flux-limited samples with well understood selection
functions that allow one to evaluate the volume surveyed and thus lead
to a straightforward computation of comoving number densities.
Fig.~\ref{fig:3}  illustrates the sky coverage of several X-ray 
surveys carried out over the last two decades.
Completeness is an important quantity in observational cosmology. A
well defined and complete sample is designed to detect all objects
with luminosity (or any other cluster quantity) above a given value
and within a given redshift, and thus it can be reliably used for
cosmological studies.

\medskip\noindent However, the most important cluster parameter, its
mass, is not direcly observable. So observers generally proceed by
using some other observable like X-ray luminosity or temperature as a
surrogate for cluster mass and linking that observable with mass
through a simple scaling relation. Numerical simulations of cluster
formation indicate that these relations can be quite accurate (e.g.,
\cite{evr96} and \cite{bn98} show in simulations without radiative
cooling or star formation, that cluster temperature tracks cluster
mass to within about 15\%; see among others also \cite{muan02,ett04a,
kra05} for simulations with cooling and star formation).  Several
proxies of the total cluster mass have been proposed based on cluster
observables such as galaxy velocity dispersion \cite{gir98}, optical
light \cite{pos96,gy05}, mass of the ICM (many papers by many
authors), Sunyaev-Ze'ldovich decrement \cite{car02}, gravitational
lensing \cite{fm94,twv90,ks93} (see \cite{voit05a} for description of
all these methods). Two easy to obtain X-ray observables are X-ray
luminosity and X-ray gas temperature, which are both found to correlate
more tightly with the cluster virial mass than other cluster
properties like for instance optical richness.

\medskip\noindent The cluster X-ray luminosity (L$_{X}$) is the most
straightforward mass indicator to measure observationally since a
minimum number of X-ray photons is required. However, since most of
the luminosity comes from the dense central region of the clusters
(the radius of the core is much smaller than the virial radius),
L$_{X}$ is the least accurate of all proposed X-ray indicators for the
total mass given the large scatter and deviations of the slope of the
luminosity-mass, L$_{X}$-M, relation \cite{rb02} from self-similar
model predictions\footnote{Those ICM models whose physics is based on
the assumption that only gravity determines the thermodynamical
properties of the hot diffuse gas are called self-similar models
\cite{kai86}. In such models clusters of different sizes are expected
to be scaled version of each other since gravity does not have a
preferred scale.}.  One way to calibrate the L$_{X}$-M relation is to
combine the M--T$_{X}$ relation (whose scatter has been found to be
considerably smaller e.g., \cite{fin01,arn05,vik06}) with the observed
L$_{X}$-T$_{X}$ relation. Observational studies have found that the
slope of the L$_{X}$-T$_{X}$ relation is steeper than self-similar
predictions (e.g., \cite{ma98, ae99}), and the entropy in cluster
cores is higher than predicted (e.g., \cite{pon03}) indicating 
important non-gravitational effects (such as cooling, mergers,
etc) on the energy budget of clusters.  One has to pay attention also
to the evolution of the mass-observable relations.  For instance,
Branchesi and collaborators \cite{bra07a,bra07b} find a significant
evolution in the $L_{\rm X}$-T of a sample of 40 archival {\em
Chandra} and XMM clusters, similar or stronger than the self-similar
model, from z = 0 to $z \le 0.3$, followed by a much weaker, if any,
evolution at higher redshifts (see also \cite{mau08}). The higher-z
weaker evolution seems compatible with an increasing importance at
high redshift of non-gravitational effects in the structure formation
process (e.g. \cite{voit05a,voit05b}).

\medskip\noindent 
The X-ray temperature of the ICM \cite{ha91,ob92,hen04} is another
common indicator for mass. The  X-ray temperature is closely related to
the depth of a cluster potenial well and can be observed with current
X-ray detectors up to z$\sim$1 and beyond. 
Under the assumptions of hydrostatic equilibrium and isothermality
(simplifying assumptions which are not necessarily true in reality),
one can derive the total mass in X-rays by knowing the baryon density
from the X-ray surface brightness and the temperature of the hot
gas. These two quantities are readily available today with the
detectors onboard {\em Chandra} and XMM-{\em Newton} satellites which
can measure both simultaneously. The masses obtained in this way are
very close to those obtained through the virial theorem namely
T$\propto$ M$^{2/3}$. It is worth mentioning here that the very accurate 
temperature profiles out to large radii now provided by the current X-ray 
telescopes have actually allowed Vikhlinin and collaborators \cite{vik06} 
to relax the assumption of isothermality. They have used the best
available {\em Chandra} observations for thirteen low-redshift clusters 
and made direct hydrostatic mass estimates from the gas temperature profiles.

\medskip\noindent In the recent past several authors have used the
cluster baryon mass as a proxy for the total mass thus avoiding all
the uncertainties of the M--T$_{X}$ and M$_{tot}$--L$_{X}$ relations
\cite{vik03,voe04,ett03,all08}. The advantage is that it can be
measured from X-ray imaging alone and is a robust and complementary
indicator to the others for constraining cosmological parameters. An
additional recently proposed \cite{kvd06} mass indicator is defined
as the product of the X-ray derived gas mass and the averge
temperature, $Y_{X}= M_g T_X$, that strongly correlates with cluster
mass with only 5 -- 8\% intrinsic scatter.  However, non-gravitational
processes can potentially alter the mass-temperature relation, the
baryon-to-dark-matter ratio of clusters, and the redshift evolution of
both these quantities. Maughan 2007 \cite{mau07} followed up on this
and found from the $L_{X}-Y_{X}$ relation for 115 {\em Chandra}
clusters that the X-ray luminosity is a robust, low-scatter mass proxy.

\begin{figure}
\sidecaption
\includegraphics[scale=.75]{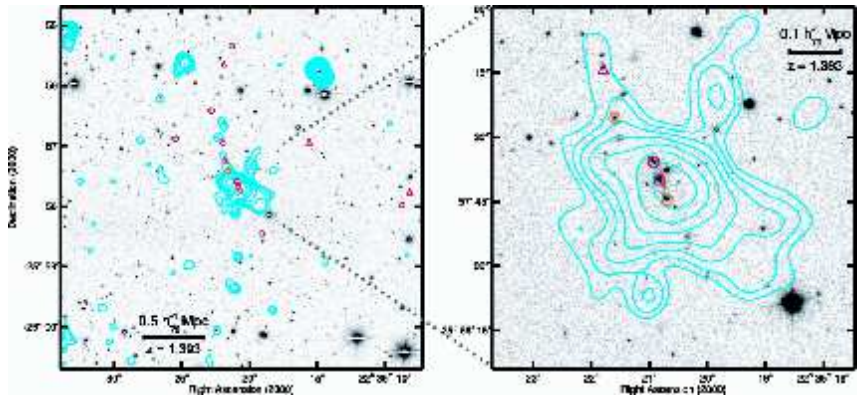}
\caption{Galaxy cluster XMMU J2235.32557, at z=1.393.  
Left: VLT FORS2 R-band image overlaid with X-ray contours from a 
45 ks XMM-{\em Newton} observation.  Right: VLT ISAAC Ks image overlaid with 
the same X-ray contours. Spectroscopically confirmed members 
(1.38$<$z$<$1.4) are indicated as circles or triangles. From Mullis 
et al. 2005.}
\label{fig:4}
\end{figure}

\medskip\noindent To wrap up this part I would say that X-ray is a
fundamental band to identify and characterize galaxy clusters. 
Current X-ray telescopes show us the very detailed fine structure of
cluster emission up to distant redshifts (see for instance
Fig~\ref{fig:4} for an XMM image of one of the most distant
serendipitously selected X-ray clusters at z=1.4, \cite{mul05})
something unthinkable until a decade ago. The many X-ray surveys from
previous missions, either serendipitous or all-sky surveys, have been
demonstrated to be promising tools for the characterization of the
properties of galaxy clusters.  I believe that we can get even more
stringent constraints on cosmological parameters as more sensitive and
statistically significant surveys made with the current telescopes
become available.

\section{Clusters of Galaxies as Cosmological Tools}
\label{sec:3}
\begin{figure}[b]
\sidecaption[b]
\includegraphics[scale=.37]{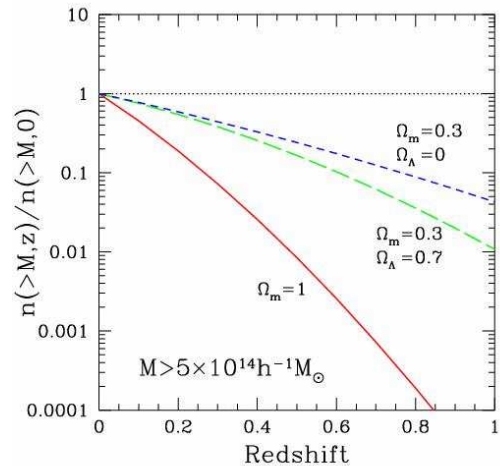}
\caption{Evolution of n($>$M, z) for M $>$ 5 $\times$ 10$^{14}$ h$^{-1}$
M\sun for three cosmologies (solid line, $\Omega_{m}$ = 1; long-dashed 
line, $\Omega_{m}$ = 0.3, $\Omega_{\Lambda}$ = 0.7; 
short-dashed line, $\Omega_{m}$ = 0.3, $\Omega_{\Lambda}$ = 0) with  
$\sigma_{8}$ = 0.5 for the $\Omega_{m}$ = 1 case 
and $\sigma_{8}$ = 0.8 for the low-density
models. From Rosati, Borgani \& Norman, 2002.}
\label{fig:5}    
\end{figure}
Clusters of galaxies are the highest peaks in a cosmic terrain driven
by gravitational clustering and represent the largest scale of fully
collapsed structures in the Universe \cite{pee93,pea99}.  Thus they
offer a unique insight into the formation of structures and into the
parameters governing their evolution.  The internal mix of components
within clusters, as well as the space density and temperature
distribution function of the most distant and massive clusters, can be
used to determine fundamental cosmological parameters. Other clusters
measurements useful for cosmological studies include the power
spectrum of the three-dimensional distribution of clusters, and the
baryon fraction and its evolution. These studies have been carried out
by a number of authors over the years.  Among them see for instance:
\cite{hen97,eke98,bor99,bor01,ros02,ett03,pier03,
hen04,vik03,voe04,all08} which is not a complete list.  These
works have used the mass function as given by Press-Schecter
\cite{ps74} or by Sheth \& Tormen \cite{st99} or by Jenkins
\cite{jen01}.  The values of the mean mass density, $\Omega_{m}$, and
dark energy density, $\Omega_{\Lambda}$, of the universe are
fundamental data for cosmological theories. These quantities are
conveniently parameterized in terms of the critical density,
$\rho_{0}$ = 3H$_{0}^{2}$/(8$\pi$G) (here $H_{0}$ = 100 h km s$^{-1}$
Mpc$^{-1}$ is the Hubble constant, G is the gravitational constant and
h is the value of the Hubble parameter in units of 100 km s$^{-1}$
Mpc$^{-1}$). The growth rate of the density perturbations depends
primarily on $\Omega_{m}$ and, to a lesser extent, on
$\Omega_{\Lambda}$ at least out to z$\approx$1 where we can study
clusters observationally.  The abundance of rich clusters of galaxies
is extremely sensitive to the amplitude of the mass density
fluctuations on a scale of 8 h$^{-1}$ Mpc, or $\sigma_8$ \footnote{
$\sigma_8$ describes the amount of structure in the Universe and is
represented as the rms matter fluctuations in spheres of 8 h$^{-1}$
Mpc.}, while the evolution of the abundance is extremely sensitive to
$\Omega_{m}$ and to a lesser extent to $\Omega_{\Lambda}$.  An
additional parameter is the dark energy parameter {\em w}
\cite{cds98}, the ratio between the pressure and energy density in the
equation of state of the dark energy component. \footnote{If
\textit{w}=-1 then the dark energy is the cosmological constant, if
-1$<$\textit{w}$<$0 then it is called ``quintessence'' or Q component
\cite{cds98}.}  The value of \textit{w} is less constrained by
clusters.

\begin{figure}[t]
\sidecaption[t]
\includegraphics[scale=.36]{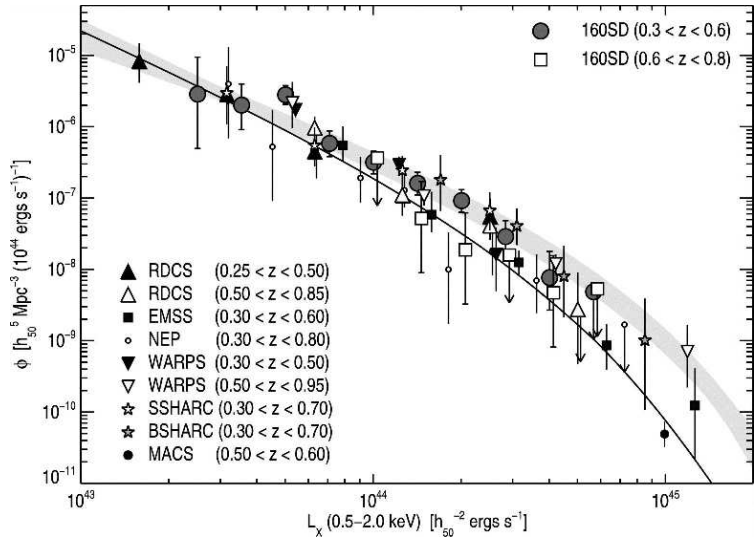}
\caption{Compilation of high-redshift XLFs as measured by eight
independent X-ray flux limited surveys. The shaded region delineates 
the regime of the local XLF, whereas the solid line is an evolving 
model XLF. From Mullis et al. 2004.}
\label{fig:6}    
\end{figure}

\medskip\noindent Fig~\ref{fig:5} shows the sensitivity of the
cluster mass function to cosmological models.  Both the X-ray
luminosity function (XLF, the number density of galaxy clusters having
a given X-ray luminosity) and the X-ray temperature function (XTF, the
number density of galaxy clusters having a given temperature) for both
nearby and distant clusters have been used as a proxy for the mass
function by a number of authors.  When only local cluster data are
used, there is a degeneracy between $\sigma_8$ and $\Omega_{m}$.
See discussion in \cite{bor99,bor01} on how the resulting constraints 
on the $\sigma_8$--$\Omega_{m}$ plane vary by changing the parameters 
which define the M-L$_{X}$ relation.
To break this degeneracy one can either use the evolution of the XLF with
redshift, or consider measurements at other spatial scales, such as
the fluctuations in the cosmic microwave background (CMB) with appropriate
assumptions.  Many X-ray surveys have shown that the comoving number
density of clusters at a given luminosity from z$\sim$0.8 to the
present changes very little for L$_{X}\leq10^{44}$.  Evolution is
seen only for clusters with L$_{X}\geq10^{45}$ (see among others,
e.g. \cite{gio90b,ros02,mul04} and Fig.~\ref{fig:6} for a
compilation of high-redshift XLFs which highlights evolution).  
The situation becomes worse when one wants to
investigate the dark energy parameter {\em w}. In that case
investigators combine constraints from both Supernovae and clusters,
or weak lensing, the cosmic microwave background and clusters to 
improve the constraints.

\begin{figure}
\sidecaption
\includegraphics[scale=.42]{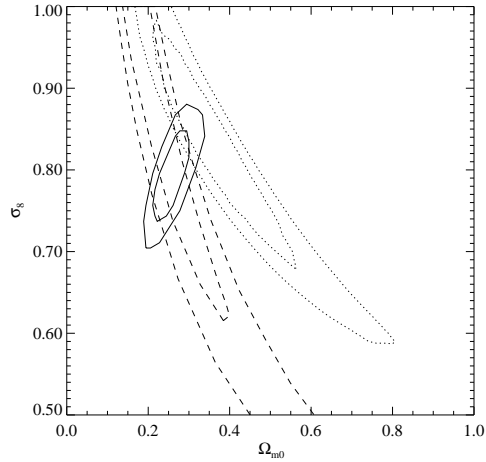}
\caption{68\% and 95\% confidence contours for two parameters for
three analyses: WMAP5 data (solid line, \cite{dun08}), cluster data 
(dotted line; Henry et al. in preparation) and weak-lensing data 
(dashed line; \cite{ben07}). All constraints intersect at 
$\sigma_8$=0.85, at $\Omega_{m0}$=0.3.} 
\label{fig:7}       
\end{figure}

\medskip\noindent The degeneracy between $\sigma_8$ and $\Omega_{m}$ may
also be broken by measuring the evolution of the cluster temperature
function. The first cosmological measurement using the evolution of
the XTF at redshift greater than zero was done by Henry in 1997
\cite{hen97} who derived $\Omega_{m}\approx 0.5\pm0.15$.  Many updates
on both theoretical and observational side followed, among others see
\cite{eke98,dv99,blanc00,bor01}. Henry 2004 \cite{hen04} measured the
X-ray temperature with ASCA of all but one cluster in the {\em
Einstein} Extended Medium-Sensitivity Survey \cite{gio90a}
high-redshift (z$\geq$0.3) sample and compared the data to a complete
sample of low-redshift clusters that also had temperature measurements
\cite{ha91}. Using constraints provided by the SNIa Hubble diagram and
the cosmic microwave background fluctuations he found that all three
bands (clusters, SN, CMB) intersect at $\Omega_{m}$ $\approx$ 0.3 and
$\Omega_{\lambda}$ $\approx$ 0.7 with the quintessence equation of
state \textit{w} = -(0.42$\pm$0.21) and $\sigma_8$=0.66$\pm$0.16.  The
last determination by the same author (Henry, in preparation) can be
considered the-state-of-the-art in the field.  Fig.~\ref{fig:7} shows
the intersect in the $\Omega_{m}-\sigma_{8}$ plane of three bands
representing three different clusters analyses. The cluster
constraints (dotted line) define a narrow band in the
$\Omega_{m}$-$\sigma_8$ plane which intersects with constraints from
the Wilkinson Microwave Anisotropy Probe Five-Year data \cite{dun08}
(solid line) (WMAP5 \footnote{WMAP mission was designed to determine
the geometry, content, and evolution of the Universe through a
full-sky map of the temperature anisotropy of the cosmic microwave
background radiation \cite{ben03a}}), and weak lensing data
\cite{ben07} (dashed line) . Allen et al.  \cite{all08} use {\em
Chandra} measurements of the X-ray mass gas fraction for 42 clusters
in the range 0.05$<$z$<$1.1 to constrain the mean matter density, the
dark energy, and dark energy parameter {\em w}.  Combining the X-ray
gas fraction f$_{gas}$ with constraints from supernova and WMAP3 
studies and for a flat cosmology they obtain a tight 
$\Omega_{m}$=0.253$\pm$0.021  and {\em w} = -0.98$\pm$0.07.  
Mantz and collaborators \cite{man08} derive a precise determination of the
dark energy equation of state combining the X-ray luminosity function data
of the most luminous clusters out to z=0.7 with supernova, WMAP3 and 
cluster gas fraction data. They find
$\Omega_{m}$ = 0.269$\pm${0.016}, $\sigma_{8}$=0.82$\pm$0.03 and {\em
w}=-1.02$\pm$0.06, in agreement with earlier galaxy cluster studies.
This demonstrates that we have already enough information from cluster
samples to also constrain the Dark Energy content of the Universe, one
of the most ambitious targets of modern cosmology.  Thus we understand
why cosmologists love to work with galaxy clusters. The reason is
simple: they are tools for precision cosmology through the evolution
of their mass function.

\begin{figure}[t]
\sidecaption[t]
\includegraphics[scale=.44]{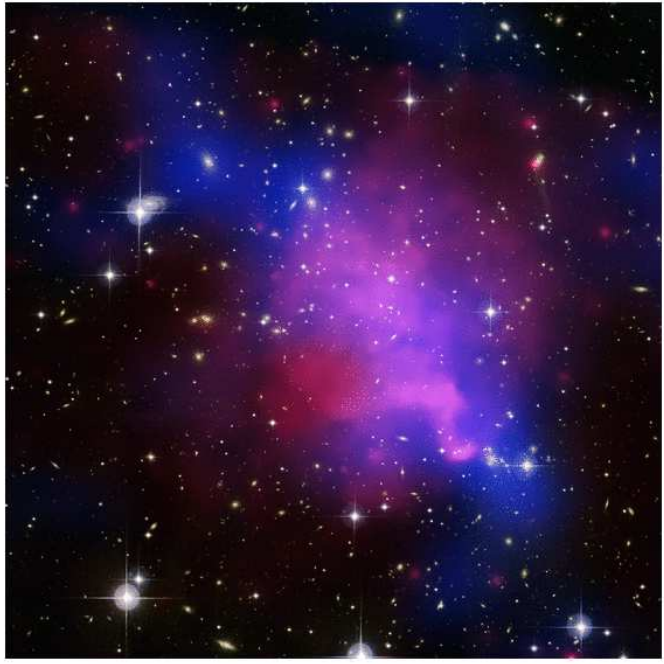}
\caption{This composite image (which appears in the {\em Chandra}
online Photo Album at 
{\em http://chandra.harvard.edu/photo/2007/a520/index.html}) 
contains three views of the cluster Abell 520. The hot gas as detected
by {\em Chandra} is colored red.  Optical data from the
Canada-France-Hawaii and Subaru telescopes show the starlight from the
individual galaxies (yellow and orange). The location of most of the
matter in the cluster (blue) was also found using these telescopes, by
means of weak gravitational lensing of the distant galaxies by the
intervening matter.  Credits: NASA/CXC/CFHT/University of
Victoria/A. Mahdavi et al. (2007). Courtesy of A. Mahadavi and CXC.}
\label{fig:8}
\end{figure}

\section{Multi-Wavelength Approach}
\label{sec:4}
A multi-wavelength approach in any branch of astronomy is of
importance since different bands highlight different properties of the
emission mechanisms or detect different components of the astronomical
objects which contribute to our understanding of their physics,
formation and evolution.  The composite image
of Abell 520 in Fig.~\ref{fig:8} highlights the
usefulness of the multi-wavelength approach to detect different
emission sources from clusters of galaxies  \cite{mah07}. The
different waveband measurements are complementary in setting more
stringent cosmological constraints. X-ray clusters can be used alone
to constrain cosmological parameters or can be combined with
independent methods (weak-lensing, CMB anisotropies,
Sunyaev-Zel'dovich, supernovae, to name a few) and different
wavelength data (optical, radio, infrared etc.).  As we have seen the
cluster mass is a parameter of great value in observational cosmology.
A combination of several, independent cluster mass estimates is likely
to provide the most accurate results.

\medskip\noindent In the optical the mass-to-light ratio or the
mass-richness relation, as well as mass estimates based on the
dynamics of member galaxies, have been used by a number of
authors with some success, \cite{gir98,gir00,biv06,gla07}.  
I would like to mention here some of the optical cluster surveys which
have overcome the problem of projection effects. I am refererring to
the work of Gladders \& Yee \cite{gy05} (the Red-Sequence Cluster 
Survey) who demonstrated  that two filter imaging is
sufficient to perform a clean cluster search using the cluster red
sequence of early-type galaxies, even when probing deeply into the
mass function.   Zaritsky and collaborators (see
Las Campanas Distant Cluster Survey \cite{gon01}) adopted
a different method where clusters are detected as positive surface 
brightness fluctuations in the background sky.

\medskip\noindent An unquestionable unique tool to study the matter
distribution of the Universe is the use of the weak gravitational
lensing of distant galaxies by intervening matter.  We have seen in
the previous section how the use of weak-lensing coupled with CMB and
X-ray data has led to much more stringent constraints on $\Omega_{m}$
and $\sigma_8$. Weak lensing has benefitted from the excellent optical
surveys currently available with multi-color data and superb image
quality over wide areas \cite{clo06,ben07,fu08}. The larger areas
enable the measurement of the lensing signal out to much larger radii,
thus improving the reliability of the results \cite{hoek07}.  See
Hoekstra and collaborators \cite{hoek02} for a review on weak gravitational
lensing.

\medskip\noindent In the radio band the pioneering work of Feretti and
collaborators, i.e. \cite{fer96,fer04,fer05,gov01,gov04,git07}, have
unveiled large diffuse cluster components in the ICM due to synchroton
radio emission not directly related to the cluster galaxies.  The
study of these sources (called radio halos, relics and mini-halos
according to their size, shape and location with respect to the cluster
center) is very important since they are large scale features which are
related to other cluster properties in the optical and X-ray domain,
and are thus directly connected to the cluster history and
evolution. The radio halos are indicators of cluster mergers, probes
of the ICM magnetic fields. They will eventually allow us to constrain
models of decaying/annihilating dark matter species. The radio relics
are likely tracers of shock waves during the structure formation. The
radio mini-halos are found in the center of clusters with cooling
cores and will allow us to investigate the interaction between the
relativistic plasma and the thermal plasma at the cluster centers.
The future Square Kilometer Array
\footnote{The radio telescope will have an effective collecting area
more than 30 times greater than the largest current telescope 
``Galaxy evolution, cosmology and dark energy'' is one of
five projects identified by the radio astronomy community as being  
the key science drivers for the SKA.}  (SKA) will
dramatically improve the knowledge of these sources, thanks to the
detection of new objects, and to detailed studies of their spectra and
polarized emission. See also \cite{burig04} for the contribution of SKA
to future CMB spectrum space experiments.

\noindent I would like to mention that the non-thermal component in
clusters with radio halos has been detected in the hard (above 25 keV)
X-ray band (HRX) due to inverse Compton scattering by relativistic
electrons of the cosmic microwave background photons. Fusco-Femiano
and collaborators \cite{fus99} found it in Coma, A2256 and A754, among
other clusters.  This is another manifestation of the same
relativistic electrons which emit by synchroton in the radio band. The
detection of the non-thermal HRX has enjoyed healthy debate up until
now among the different observers.
\begin{figure}
\sidecaption
\includegraphics[scale=.80]{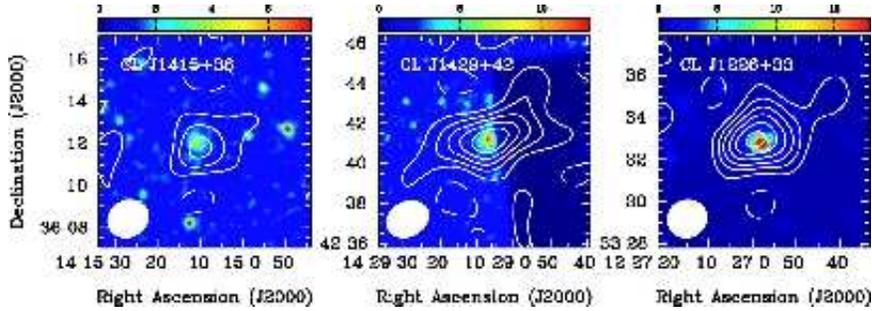}
\caption{SZE effect measurements (contours) overlaid on 
XMM-{\em Newton} X-ray images of Cl J1415.1+3612 (z=1.03), 
Cl J1429.0+4241 (z=0.92), and Cl J1226.9+3332 (z=0.89) 
(from left to right).  The SZE observations were obtained at 
30 GHz during the commissioning period of the new, eight-element 
interferometer SZA (Sunyaev-Zeldovich Array).  In each panel, the 
FWHM of the synthesized beam of the SZE observations is shown by 
the filled ellipse in the bottom left corner. 
Taken from Muchovej et al. (2007).}
\label{fig:9}
\end{figure}

\medskip\noindent Another powerful observational tool for cosmology is
the Sunyaev Zel'dovich effect or SZE (see review by
\cite{bir99,car02}), which is a distorsion in the CMB spectrum caused
by the CMB photons passing through the hot ICM and inverse Compton
scattering off the energetic electrons. The effect is insensitive to
the redshift of the clusters, thus making the method well suited for
studies of clusters at high redshift where the abundance of galaxy
clusters critically depends on the underlying cosmology. While the
thermal SZE is a function of the electron number density, n$_e$, the
X-ray emission scales as n$^{2}_{e}$. Thus clusters are more extended
when mapped in SZE than in X-rays (see Fig~\ref{fig:9} for
X-ray and SZE maps of three distant clusters).  The different
dependence on the gas density enables a determination of the direct
distance to the galaxy cluster which is independent of the
extragalactic distance ladder, up to high-z clusters.  The great merit
of SZE is that combined with other observational diagnostics of
clusters (X-ray emission, weak and strong lensing, galaxy velocity
dipersion measurements) can provide a measure of the basic
cosmological parameters like for instance the Hubble constant.  Recently
Bonamente et al. \cite{bon06} used 38 clusters with {\em Chandra} and SZE
data to find a value for H$_{0}$ = 76.9$^{+3.9}_{-3.4}$ km s$^{-1}$
Mpc$^{-1}$ (68\% confidence level) with $\Omega_m$=0.3 and
$\Omega_{\lambda}$=0.7 cosmology in agreement with result obtained by
the Hubble Space Telescope for clusters at low redshift, and with the
$\Lambda$CDM concordance model\footnote{According to the $\Lambda$CDM
model the Universe is spatially flat, homogeneous and isotropic on
large scales, composed of radiation, ordinary matter (electrons,
protons, neutrons, and neutrinos), nonbaryonic cold dark matter, and
dark energy and with adiabatic initial conditions of the density
fluctuations. The direct confirmation of this theory was the detection
of the acoustic Doppler peak structure in the CMB angular power
spectrum.  A wide range of astronomical datasets are consistent with
the predictions of the $\Lambda$CDM model with its parameters fitted
to the WMAP data. These data range from large scale structure galaxy
surveys, supernovae luminosity distance, Lyman-$\alpha$ forest, weak
and strong lensing etc. \cite{ben03b,spe03,spe07}.}.
 
\medskip\noindent The Cosmic Soft Excess (CSE) is a phenomen exhibited
by a fraction of clusters ($\approx 30-40\%$) in the extreme
ultraviolet or in the soft (1 keV) X-ray band \cite{lm04}. Since its
discovery the properties and origin of the CSE have been subject of
debate. CSE has been detected by the Extreme Ultraviolet Explorer and
by several X-ray telescopes including the current XMM-{\em Newton}. Both
the thermal and non-thermal interpretation on the CSE origin have been
considered and the issue is still under study.

\section{In Conclusion}
\label{sec:5}

The wealth of astrophysical information currently available on galaxy 
clusters can give us a deeper understanding of the Universe we live
in. We have entered in a promising era for cosmology with clusters.
Today scientists are adopting the multi-frequency approach to carry
out cosmological studies since each wavelength contributes a little
piece of information which makes sense once the whole puzzle is
assembled. We have come a long way since the times when astronomers 
were looking for overdensities of galaxies to discover clusters!  X-ray
observations, optical and infrared observations of the cluster member
galaxies and weak lensing of background galaxies by the deep cluster
potential are complementary probes of high-redshift clusters.  
Measurements of the SZE have been used to determine cluster
properties such as the gas and total masses, electron temperatures, as
well as to constrain the cosmological distance scale 
\cite{gre01,lar06,bon06}.  The SZE will soon be used as a new band for
detecting clusters at high redshift. SZE surveys will be a tremendous
source of new information in the near future. In particular surveys
like the South Pole Telescope\footnote{http://pole.uchicago.edu/}
(SPT; \cite{sp04}) or the Atacama
Cosmology Telescope\footnote{http://www.physics.princeton.edu/act/} 
(ACT; \cite{kos03}) will produce
catalogs of clusters unbiased in redshift. Some of the planned SZE
instrumentation is now reality. I am thinking of the
Sunyaev-Zel'dovich Array (SZA), an eight-element interferometer which
enables one to achieve high sensitivity with respect to single dish
observations even for extended low-surface brightness emission.  During
the commissioning period the SZA demonstrated that it can be used to
study distant (z$\geq$1) clusters \cite{much06}.

\medskip\noindent The soon to be flown Planck satellite\footnote{ {\em
http://www.rssd.esa.int/index.php?project=Planck}} will extend our
knowledge of the CMB beyond the limits set by past and present
experiments (for instance WMAP). Planck will survey the whole sky and
will provide a large dataset of clusters expected to be at z $>$ 1.
Blind SZE surveys in the near future will discover thousands of
clusters. Since the SZE signal is independent of redshift the limit of
such surveys will be a mass limit. Such cluster surveys can be used to
determine cosmological parameters with high precision.  The Planck
mission will likely lead us to a full comprehension of the CMB
temperature anisotropies and it will be crucial as a test of the
robustness of the CDM concordance model.

\medskip\noindent The non-thermal components of clusters of galaxies
will be revealed by the future radio telescopes. When arrays like the
Square Kilometer Array, SKA\footnote{\em http://www.skatelescope.org/},
the Long Wavelength Array, LWA\footnote{\em
http://lwa.unm.edu/}, or the Low Frequency Array,
LOFAR\footnote{\em http://www.lofar.org/}, will become operational they
will reveal new radio halos especially in distant clusters. One can
then be able to compare the statistics between the observationl data
and the expectations from models of cluster and structure formation.
The combination of radio and hard X-ray data will be crucial to
measure the energy content in the form of relativistic electrons and
magnetic field in the ICM. The proposed new generation hard X-ray
telescope SIMBOL X \footnote{\em http://www.asdc.asi.it/simbol-x/}
(a jointly supported Italian-French mission with German participation) 
which will operate in the 0.5--80 keV, is expected  to reveal and map 
the non-thermal emission in clusters of galaxies.

\medskip\noindent I believe the time is certainly mature to have a new
medium-depth X-ray all-sky survey of clusters carried out with a dedicated
satellite with a good point spread function (similar or better than
XMM), optimized optics for wide-field X-ray imaging and low
background.  An all-sky survey, and its associated large sample of
clusters, would be crucial to investigate the relationship between
X-ray observables and masses. In addition many new clusters
at high redshift will be discovered. We need more objects to observe
and study.  Several ideas for such a survey have been proposed by the
scientific community to the various space agencies but none has been
approved so far. In the meantime we have to make do with the invaluable
archives of both {\em Chandra} and XMM-{\em Newton} which are
providing interesting new results (see among others
\cite{ett04b,voe04,mul05,vik06,bra07b,bra07a}) and with ongoing
X-ray cluster surveys like the X-ray Cluster Survey (XCS;
\cite{rom01}) or the XMM-Large Scale Structure survey (XMM-LSS;
\cite{pie07}). The first will produce a catalog of several thousand
serendipitously-detected clusters in over 500 deg$^{2}$ to beyond
z$=$1. See \cite{sah08} for a recent paper forecasting the constraints
on the values of $\Omega_{m}$, $\sigma_{8}$, and cluster scaling
relations parameters expected from the XCS survey. The second survey,
the XMM-LSS, has recently produced a combined analysis of weak lensing
and X-ray blind surveys \cite{berg08}.  Meanwhile the continuing
program of {\em Chandra} and XMM observations will contribute to
increase the cluster statistics. The Planck satellite will provide new
large datsets of clusters identified through the SZ effect.  These
will be new targets for the future X-ray observatories like the NASA
mission Constellation-X
\footnote{{\em http://constellation.gsfc.nasa.gov/}} and 
the ESA mission XEUS 
\footnote{{\em http://www.rssd.esa.int/index.php?project=XEUS}}
that will allow us to carry out more precise studies on the 
nature and content of the Dark Matter and Dark Energy of the Universe.

\section{Further Reading}
\label{sec:6}
Additional articles on Cosmology with Clusters not appearing
in the Reference section are listed  below:
\begin{itemize}

\item Borgani, S. \&  Guzzo, L.:
X-ray Clusters of Galaxies as Tracers of Structure in the Universe, 
Nature, \textbf{409}, 39--4  (2001)

\item  Borgani, S.: Cosmology with Clusters of Galaxies
eprint arXiv:astro-ph/0605575, to appear in ``Lecture Notes in Physics'' 

\item Peebles, P.J.E. \& Ratra, B.: The Cosmological Constant and Dark 
Energy,  Reviews of Modern Physics, \textbf{75}, page 569--606 (2003)

\item Tozzi, P.: Cosmological parameters from Galaxy Clusters: An
Introduction, Lecture Notes in Physics, \textbf{720}, 125--156 (2007)
\end{itemize}

\begin{acknowledgement}
I warmly thank the hospitality of the Institute for Astronomy of the
University of Hawai'i where most of this work was completed during the
spring of 2008. I enjoyed fruitful discussions with M. Branchesi,
L. Feretti, R. Gal, F. Govoni, P. Henry, M. Murgia and B. Tully. 
I acknowledge partial financial support from contracts ASI-INAF
I/023/05/0, 088/06/0 and DA-030.

\end{acknowledgement}
%
%

\input{referenc}

\end{document}

%% file: referenc.tex
%
%
%

%% file: gioia.bbl
\begin{thebibliography}{99.}

\bibitem{a58} 
Abell, G.O.: The Distribution of Rich Clusters of Galaxies.
ApJS, \textbf{3}, 211-288 (1958)
\bibitem{all08}
Allen, S.W., Rapetti, D. A., Schmidt, R.W., Ebeling, H., Morris, R.G., \&
Fabian, A.C.:  Improved constraints on dark energy from Chandra X-ray 
observations of the largest relaxed galaxy clusters, MNRAS, \textbf{383},
879--896 (2008)
\bibitem{ae99}
Arnaud, M. \&  Evrard, A.E.: The L$_{X}$-T relation and intracluster gas 
fractions of X-ray clusters, MNRAS, \textbf{305}, 631--640 (1999)
\bibitem{arn05}
Arnaud, M., Pointecouteau, E., \&  Pratt, G.W.: The structural and scaling 
properties of nearby galaxy clusters. II. The M-T relation, A\&A, 
\textbf{441}, 893--903 (2005)
\bibitem{ben07}
Benjamin, J., Heymans, C., Semboloni, E., van Waerbeke, L., Hoekstra, H.,
Erben, T., Gladders, M.D., Hetterscheidt, M., Mellier, Y., \&  Yee, 
H.K.C.: Cosmological constraints from the 100-deg2 weak-lensing survey,
MNRAS, \textbf{381}, 702--712 (2007)
\bibitem{bach03}
Bacchi, M., Feretti, L., Giovannini, G., \& Govoni, F. A\&A, \textbf{400}, 
465--476 (2003)
\bibitem{ben03a}
Bennett, C.L., Bay, M., Halpern, M., Hinshaw, G., Jackson, C.,
Jarosik, N., Kogut, A., Limon, M., Meyer, S.S., Page, L., Spergel,
D.N., Tucker, G.S., Wilkinson, D.T., Wollack, E., \& Wright, E.L.:
The Microwave Anisotropy Probe Mission, ApJ, \textbf{583}, 1--23 (2003a)
\bibitem{ben03b}
Bennett, C.L., Halpern, M., Hinshaw, G., Jarosik, N., Kogut, A.,
Limon, M., Meyer, S.S., Page, L., Spergel, D.N., Tucker, G.S.,
Wollack, E., Wright, E.L., Barnes, C., Greason, M.R., Hill, R.S.,
Komatsu, E., Nolta, M.R., Odegard, N., Peiris, H.V., Verde, L., \&
Weiland, J.L.: First-Year Wilkinson Microwave Anisotropy Probe (WMAP)
Observations: Preliminary Maps and Basic Results, ApJS, \textbf{148},
1--27 (2003b)
\bibitem{berg08}
Berg, J., Pacaud, F., R\'efr\'egier, A., Massey, R., Pierre, M., 
Amara, A., Birkinshaw, M., Paulin-Henriksson, S., Smith, G.P., \& Willis, 
J.: Combined analysis of weak lensing and X-ray blind surveys, 
arXiv:0712.3293, MNRAS, in press (2008)
\bibitem{bir99}
Birkinshaw, M.: The Sunyaev-Zel'dovich effect,  PhR, \textbf{310},
97--195 (1999)
\bibitem{biv06}
Biviano, A., Murante, G., Borgani, S., Diaferio, A., Dolag, K., \&
Girardi, M.: On the efficiency and reliability of cluster mass
estimates based on member galaxies, A\&A, \textbf{456}, 23--36 (2006)
\bibitem{blanc00}
Blanchard, A., Sadat, R., Bartlett, J.G., \& Le Dour, M.: A new local 
temperature distribution function for X-ray clusters: cosmological 
applications, A\&A, \textbf{362}, 809--824 (2000)
\bibitem{bon06}
Bonamente, M., Joy, M.K., LaRoque, S.J., Carlstrom, J.E., Reese,
E.D., \& Dawson, K.S.: Determination of the Cosmic Distance Scale
from Sunyaev-Zeldovich Effect and Chandra X-Ray Measurements of
High-Redshift Galaxy Clusters, ApJ, \textbf{647}, 25--54 (2006)
\bibitem{bor99}
Borgani, S., Rosati, P., Tozzi, P., \&  Norman, C.: Cosmological Constraints 
from the ROSAT Deep Cluster Survey, ApJ, \textbf{517}, 40--53 (1999)
\bibitem{bor01}
Borgani, S., Rosati, P., Tozzi, P., Stanford, S.A., Eisenhardt, P.R., 
Lidman, C., Holden, B., Della Ceca, R.,  Norman, C., \& Squires, G.:
Measuring $\Omega_{0}$ with the  ROSAT Deep Cluster Survey, ApJ, 
\textbf{561}, 13--21 (2001)
\bibitem{bra07a}
Branchesi, M., Gioia, I.M., Fanti, C., \&  Fanti, R.:  High redshift X-ray 
galaxy clusters. II. The LX-T relationship revisited, A\&A, \textbf{472}, 
739--748 (2007a)
\bibitem{bra07b}
Branchesi, M., Gioia, I.M., Fanti, C., \&  Fanti, R.: High redshift 
X-ray galaxy clusters. I. The impact of point sources on the cluster 
properties, A\&A, \textbf{472} 727--737 (2007b)
\bibitem{bn98}
Bryan, G.L. \& Norman, M.L.:
Statistical Properties of X-Ray Clusters: Analytic and Numerical Comparisons
ApJ,  \textbf{495}, 80--99 (1998)
\bibitem{bur07}
Burenin, R.A., Vikhlinin, A., Hornstrup, A., Ebeling, H., Quintana, H., 
Mescheryakov, A.: The 400 Square Degree ROSAT PSPC Galaxy Cluster Survey: 
Catalog and Statistical Calibration, ApJS, \textbf{172}, 561--582 (2007)
\bibitem{burig04}
Burigana, C., De Zotti, G., \& Feretti, L.: Sunyaev-Zel'dovich effetcs,
free-free emission, and imprints on the cosmic microwve background, 
NewAR,  \textbf{48}, 1107-1117 (2004)
\bibitem{cds98}
Caldwell, R.R., Dave, R. \& Steinhardt, P.J.: Cosmological Imprint of an
Energy Component with General Equation of State, PhRvL, \textbf{80}, 
1582--1585 (1998)
\bibitem{car02}
Carlstrom, J.E., Holder, G.P., \& Reese, E.D.:
Cosmology with the Sunyaev-Zel'dovich Effect, ARA\&A, \textbf{40}, 
643--680 (2002)
\bibitem{clo06} 
Clowe, D., Schneider, P., Aragn-Salamanca, A., Bremer, M., De Lucia,
G., Halliday, C., Jablonka, P., Milvang-Jensen, B., Pell, R.,
Poggianti, B., Rudnick, G., Saglia, R., Simard, L., White, S.,
Zaritsky, D.; Weak lensing mass reconstructions of the ESO Distant
Cluster Survey, A\&A, \textbf{451}, 395--408 (2006)
\bibitem{dv99}
Donahue, M., Voit, G.M.:  Omega$_{m}$ from the Temperature-Redshift 
Distribution of EMSS Clusters of Galaxies, ApJL, \textbf{23}, 137--140 
(1999)
\bibitem{dun08} 
Dunkley, J., Komatsu, E., Nolta, M.R., Spergel, D.N., Larson, D.,
Hinshaw, G., Page, L., Bennett, C.L., Gold, B., Jarosik, N., Weiland,
J.L., Halpern, M., Hill, R.S., Kogut, A., Limon, M., Meyer, S.S.,
Tucker, G.S., Wollack, E., Wright, E.L.: Five-Year Wilkinson Microwave
Anisotropy Probe (WMAP) Observations: Likelihoods and Parameters from
the WMAP data, arXiv:0803.0586 (2008)
\bibitem{eke98}
Eke, V.R., Cole, S., Frenk, C.S, \&  Henry, J.P.:
Measuring $\Omega_{0}$ using cluster evolution, MNRAS, \textbf{298}, 
1145--1158 (1998)
\bibitem{ett03}
Ettori, S., Tozzi, P., \&  Rosati, P.: Constraining the cosmological 
parameters with the gas mass fraction in local and z$>$0.7 galaxy 
clusters, A\&A,  \textbf{398}, 879--890 (2003)
\bibitem{ett04a} 
Ettori, S., Borgani, S., Moscardini, L., Murante, G., Tozzi, P., 
Diaferio, A., Dolag, K., Springel, V., Tormen, G.,
Tornatore, L.:  Evolution at z$\geq$0.5 of the X-ray properties of
simulated galaxy clusters: comparison with observational constraints,
MNRAS, \textbf{354}, 111--122 (2004a)
\bibitem{ett04b}
Ettori, S., Tozzi, P., Borgani, S., \&  Rosati, P.: Scaling laws in X-ray 
galaxy clusters at redshift between 0.4 and 1.3, A\&A. \textbf{417},
13--27 (2004b)
\bibitem{evr96}
Evrard, A.E., Metzler, C.A., \&  Navarro, J.F.: Mass Estimates of 
X-Ray Clusters, ApJ, \textbf{469}, 494--507 (1996)
\bibitem{fer96}
Feretti, L. \&  Giovannini, G.: Diffuse Cluster Radio Sources, IAUS,
\textbf{175}, 333--338 (1996)
\bibitem{fer97}
Feretti, L., B\"ohringer, H., Giovannini, G., \&  Neumann, D.: 
The radio and X-ray properties of Abell 2255, A\&A, \textbf{317},
432--440 (1997)
\bibitem{fer04}
Feretti, L., Burigana, C., \&  Ensslin, T.A.: Diffuse radio emission 
from the intracluster medium, NewAR,  \textbf{48}, 1137-1144 (2004)
\bibitem{fer05}
Feretti, L.: Non-thermal emission from the intracluster medium
AdSpR, \textbf{36}, 729--737 (2005)
\bibitem{fin01}
Finoguenov, A., Reiprich, T.H., \& B\"ohringer, H.: 
Details of the mass-temperature relation for clusters of galaxies, A\&A, 
\textbf{368}, 749--759 (2001)
\bibitem{fm94}
Fort, B. \&  Mellier, Y.: Arc(let)s in clusters of galaxies, A\&ARev,
\textbf{5}, 239--292 (1994)
\bibitem{fu08}
Fu, L., Semboloni, E., Hoekstra, H., Kilbinger, M., van Waerbeke, L.,
Tereno, I., Mellier, Y., Heymans, C., Coupon, J., Benabed, K.,
Benjamin, J., Bertin, E., Dore', O., Hudson, M. J., Ilbert, O., Maoli,
R., Marmo, C., McCracken, H.J., \& Menard, B.: Very weak lensing in the
CFHTLS wide: cosmology from cosmic shear in the linear regime, A\&A,
\textbf{479}, 9--25 (2008)
\bibitem{fus99} 
Fusco-Femiano, R., Dal Fiume, D., Feretti, L., Giovannini, G., Grandi,
P., Matt, G., Molendi, S., \& Santangelo, A..: Hard X-Ray Radiation in
the Coma Cluster Spectrum, ApJL, \textbf{513}, 21--24 (1999)
\bibitem{gio90a} 
Gioia, I.M., Maccacaro, T., Schild, R.E.,  Wolter, A., Stocke, J.T., 
Morris, S.L., \&  Henry, J.P.: The Extended Medium Sensitivity Survey.I -
X-ray data and analysis, ApJS, \textbf{72}, 567--619 (1990a)
\bibitem{gio90b} 
Gioia, I.M., Henry, J.P., Maccacaro, T., Morris,,S.L., Stocke, J.T., 
\& Wolter, A.: The Extended Medium Sensitivity Survey distant cluster 
sample - X-ray cosmological evolution, ApJL, \textbf{356}, 35--38 (1990b)
\bibitem{gio03}
Gioia, I.M., Henry, J.P., Mullis, C.R., B\"ohringer, H., Briel,
U.G., Voges, W., \&  Huchra, J.P.:  The ROSAT North Ecliptic Pole
Survey: the Optical Identifications, ApJS, \textbf{149}, 29--51 (2003)
\bibitem{gg99} 
Giovannini, G., Tordi, M., \&  Feretti, L.: Radio halo and relic 
candidates from the NRAO VLA Sky Survey, NewAR, \textbf{4}, 141--155
 (1999)
\bibitem{gir98}
Girardi, M., Giuricin, G., Mardirossian, F., Mezzetti, M., \& Boschin, W.:
Optical Mass Estimates of Galaxy Clusters, ApJ, \textbf{505}, 74--95 
(1998)
\bibitem{gir00}
Girardi, M., Borgani, S. Giuricin, G., Mardirossian, F., \& Mezzetti, M.:
Optical Luminosities and Mass-to-Light Ratios of Nearby Galaxy Clusters,
ApJ, \textbf{530}, 62--79 (2000)
\bibitem{git07}
Gitti, M., Ferrari, C., Domainko, W., Feretti, L., \& Schindler, S.: 
Discovery of diffuse radio emission at the center of the most
X-ray-luminous cluster RX J1347.5-1145, A\&AL, \textbf{470}, 25--28
(2007)
\bibitem{gy05}
Gladders, M.D. \& Yee, H.K.C.: The Red-Sequence Cluster Survey. I. The
Survey and Cluster Catalogs for Patches RCS 0926+37 and RCS 1327+29,
ApJS, \textbf{157}, 1--29 (2005)
\bibitem{gla07}
Gladders, M.D., Yee, H.K.C., Majumdar, S., Barrientos, L.F., Hoekstra,
H., Hall, P.B., \& Infante, L.: Cosmological Constraints from the 
Red-Sequence Cluster Survey, ApJ, \textbf{655}, 128--134 (2007)
\bibitem{gon01}
Gonzalez, A.H., Zaritsky, D., Dalcanton, J.J., Nelson, A.:
The Las Campanas Distant Cluster Survey: The Catalog, ApJS, \textbf{137}, 
117-138 (2001)
\bibitem{gov01}
Govoni, F., Feretti, L., Giovannini, G., B\"ohringer, H., Reiprich, T.H.,
\& Murgia, M.: Radio and X-ray diffuse emission in six clusters 
of galaxies, A\&A, \textbf{376}, 803--819 (2001)
\bibitem{gov04}
Govoni, F., Markevitch, M., Vikhlinin, A., Van Speybroeck, L.,  Feretti, 
L., \& Giovannini, G.: Chandra Temperature Maps for Galaxy Clusters with 
Radio Halos, ApJ, \textbf{605}, 695--708, (2004)
\bibitem{gov05}
Govoni, F., Murgia, M., Feretti, L., Giovannini, G., Dallacasa, D., 
Taylor, G. B.,: A2255: The first detection of filamentary polarized 
emission in a radio halo, A\&A, \textbf{430}, L5--L8 (2005) 
\bibitem{gre01}
Grego, L., Carlstrom, J.E., Reese, E.D., Holder, G.P., Holzappel, W.L.,
Marshall, K.J., Mohr, J.J., \& Pate, S.: Galaxy Cluster Gas Mass
Fractions from Sunyaev-Zel'dovich effect measurements: Constraints on
$\Omega_{M}$, ApJ, \textbf{552}, 2--14 (2001)
\bibitem{ha91}
Henry, J.P. \& Arnaud, K.A.: A measurement of the mass fluctuation 
spectrum from the cluster X-ray temperature function, ApJ,  \textbf{372},
410--418 (1991)
\bibitem{hen92}
Henry, J.P., Gioia, I.M., Maccacaro, T., Morris, S.L., Stocke, J.T.
\&  Wolter, A.:The Extended Medium Sensitivity Survey Distant Cluster 
Sample - X-ray Data and Interpretation of the Luminosity Evolution, ApJ,
\textbf{386}, 408--419 (1992)
\bibitem{hen97}
Henry, J.P.: A Measurement of the Density Parameter Derived from the
Evolution of Cluster X-Ray Temperatures, ApJL, \textbf{489}, 1--5 (1997)
\bibitem{hen04}
Henry, J.P.:  X-Ray Temperatures for the Extended Medium-Sensitivity 
Survey High-Redshift Cluster Sample: Constraints on Cosmology and the Dark 
Energy Equation of State, ApJ, \textbf{609}, 603--616 (2004)
\bibitem{hoek02}
Hoekstra, H., Yee, H..K.C. \& Gladders, M.D.: Current status of 
weak gravitational lensing, NewAR, \textbf{46}, 767--781 (2002)
\bibitem{hoek07}
Hoekstra, H.: A comparison of weak-lensing masses and X-ray properties
of galaxy clusters, MNRAS, \textbf{379}, 317--330 (2007)
\bibitem{jen01}
Jenkins, A., Frenk, C.S., White, S.D.M., Colberg, J.M., Cole, S., 
Evrard, A.E., Couchman, H.M.P., \& Yoshida, N.: 
The mass function of dark matter haloes, MNRAS, \textbf{321}, 372--384
(2001)
\bibitem{kai86}
Kaiser, N.: Evolution and clustering of rich clusters, MNRAS, \textbf{222}, 
323--345 (1986)
\bibitem{ks93}
Kaiser, N. \& Squires, G.:  Mapping the dark matter with weak 
gravitational lensing, ApJ, \textbf{404}, 441--450 (1993)
\bibitem{kra05}
Kravtsov, A.V., Nagai, D., \& Vikhlinin, A.: Effects of Cooling and
Star Formation on the Baryon Fractions in Clusters, ApJ, \textbf{625},
588--598 (2005)
\bibitem{kos03}
Kosowsky, A.: The Atacama Cosmology Telescope, NewAr, \textbf{47}, 939--943
(2003)
\bibitem{kvd06}
Kravtsov, A.V., Vikhlinin, A., \& Nagai, D.: A New Robust Low-Scatter 
X-Ray Mass Indicator for Clusters of Galaxies, ApJ, \textbf{650}, 
128--136 (2006)
\bibitem{lar06}
LaRoque, S.J., Bonamente, M., Carlstrom, J.E., Joy, M.K., Nagai, D., 
Reese, E.D., \&  Dawson, K.S.:  X-Ray and Sunyaev-Zel'dovich Effect 
Measurements of the Gas Mass Fraction in Galaxy Clusters, ApJ, \textbf{652},
917--936 (2006)
\bibitem{lm04}
Lieu, R. \& Mittaz J.: 
Soft X-ray Emission from Clusters of Galaxies and Related Phenomena,
ASSL, \textbf{309}, 2004
\bibitem{mah07}
Mahdavi, A., Hoekstra, H., Babul, A., Balam, D.D., \& Capak, P.L.:
A Dark Core in Abell 520, ApJ, \textbf{668}, 806--14 (2007)
\bibitem{man08}
Mantz, A.,  Allen, S.W.,  Ebeling, H., \&  Rapetti, D.: New constraints 
on dark energy from the observed growth of the most X-ray luminous galaxy 
clusters,  arXiv:0709.4294 (2008)
\bibitem{ma98}
Markevitch, M.:  The L$_{X}$-T Relation and Temperature Function for Nearby 
Clusters Revisited, ApJ, \textbf{504}, 27--34 (1998)
\bibitem{mau07}
Maughan, B.J.: The L$_{X}$-Y$_{X}$ Relation: Using Galaxy Cluster X-Ray 
Luminosity as a Robust, Low-Scatter Mass Proxy, ApJ, \textbf{668}, 
772--780 (2007)
\bibitem{mau08}
Maughan, B.J., Jones, C., Forman, W., \& Van Speybroeck, L.: Images,
Structural Properties, and Metal Abundances of Galaxy Clusters
Observed with Chandra ACIS-I at 0.1$<$z$<$1.3, ApJS, \textbf{174}, 
117--135 (2008)
\bibitem{muan02}
Muanwong, O., Thomas, P.A., Kay, S.T., \& Pearce, F.R.: The effect of
cooling and preheating on the X-ray properties of clusters of galaxies,
MNRAS, \textbf{336}, 527--540 (2002)
\bibitem{much06}
Muchovej, S., Mroczkowski, T., Carlstrom, J.E., Cartwright, J., Greer,
C., Hennessy, R., Loh, M., Pryke, C., Reddall, B., Runyan, M., Sharp,
M., Hawkins, D., Lamb, J.W., Woody, D., Joy, M., Leitch, E.M., Miller,
A.D.: Observations of High-Redshift X-Ray Selected Clusters with the
Sunyaev-Zel'dovich Array, ApJ, \textbf{663}, 708--716 (2006)
\bibitem{mul04}
Mullis, C.R., Vikhlinin, A., Henry, J.P., Forman, W., Gioia, I.M.,
Hornstrup, A., Jones, C., McNamara, B.R., \&  Quintana, H.:
Evolution of the Cluster X-Ray Luminosity Function, ApJ,  \textbf{607},
175--189 (2004)
\bibitem{mul05}
Mullis, C.R., Rosati, P., Lamer, G., B\"ohringer, H., Schwope, A.,
Schuecker, P., Fassbender, R.: Discovery of an X-Ray-luminous Galaxy
Cluster at z=1.4, ApJL, \textbf{623}, 85--88 (2005)
\bibitem{ob92}
Oukbir, J. \& Blanchard, A.: X-ray clusters in open universes, ApJL,
\textbf{262}, 21--24 (1992)
\bibitem{pea99}
Peacock, J.A.: Cosmological Physics, Cambridge. UK, Cambridge Univ. Press
 (1999)
\bibitem{pee93}
Peeble, P.J.E.: Physical Cosmology, Princeton, NJ, Princeton Univ.
Press (1993)
\bibitem{pier03}
Pierpaoli, E., Borgani, S., Scott, D., \& White, M.: On determining the 
cluster abundance normalization, MNRAS, \textbf{342}, 163--175 (2003)
\bibitem{pie07}
Pierre, M. and 36 coauthors: The XMM-Large Scale Structure catalogue:
X-ray sources and associated optical data, MNRAS, \textbf{382},
279--290 (2007)
\bibitem{pon03}
Ponman, T.J., Sanderson, A.J.R., \&  Finoguenov, A.: The Birmingham-CfA 
cluster scaling project - III. Entropy and similarity in galaxy systems,
MNRAS, \textbf{343}, 331-342 (2003)
\bibitem{pos96}
Postman, M., Lubin, L.M., Gunn, J.E., Oke, J.B., Hoessel, J.G., Schneider, 
D.P., \& Christensen, J.A.: The Palomar Distant Clusters Survey. I. 
The Cluster Catalog, AJ, \textbf{111}, 615--641 (1996)
\bibitem{ps74}
Press, W.H. \& Schechter, P.: Formation of Galaxies and Clusters of Galaxies 
by Self-Similar Gravitational Condensation, ApJ, \textbf{187}, 425--438 (1974)
\bibitem{rb02}
Reiprich, T.H. \& B\"ohringer, H.: The Mass Function of an X-Ray Flux-limited 
Sample of Galaxy Clusters, ApJ, \textbf{567} 716--740 (2002)
\bibitem{rom01}
Romer, A.K.,  Viana, P.T.P., Liddle, A.R., \& Mann, R G.:
Cluster Survey with XMM: Expected Catalog Properties and Scientific
Applications, ApJ, \textbf{547}, 594--608 (2001)
\bibitem{ros95}
Rosati, P., della Ceca, R., Burg, R., Norman, C., \& Giacconi, R.:
A first determination of the surface density of galaxy clusters at
very low x-ray fluxes, ApJL, \textbf{445}, 11-14 (1995)
\bibitem{ros02} 
Rosati, P., Borgani, S., \& Norman, C.: The Evolution
of X-ray Clusters of Galaxies, ARA\&A, \textbf{40}, 539--577 (2002)
\bibitem{sp04}
The South Pole Collaboration: Ruhl, J.E., Ade, P.A.R., Carlstrom, J.E.,
Cho, H.M., Crawford, T., Dobbs, M., Greer, C.H.,  Halverson, N.W.,
Holzapfel, W.L., Lantin, T.M., Lee, A.T., Leong, J.,  Leitch, E.M.,
Lu, W., Lueker, M., Mehl, J., Meyer, S.S., Mohr, J.J.,  Padin, S.,
Plagge, T., Pryke, C., Schwan, D.,  Sharp, M.K.,  Runyan, M.C.,Spieler,
H., Staniszewski, Z., \&  Stark, A.A.:  The South Pole Telescope,
SPIE, \textbf{5498}, 11--29 (2004)
\bibitem{sah08}
Sahl\'en, M., \& the XCS Collaboration, arXiv:0802.4462v1 (2008)
\bibitem{st99}
Sheth, R. K. \& Tormen, G.: Large-scale bias and the peak background 
split, MNRAS, \textbf{308}, 119--126 (1999)
\bibitem{spe03}
Spergel, D.N., Verde, L., Peiris, H.V., Komatsu, E., Nolta, M.R.,
Bennett, C.L., Halpern, M., Hinshaw, G., Jarosik, N., Kogut, A.,
Limon, M., Meyer, S.S., Page, L., Tucker, G.S., Weiland, J.L.,
Wollack, E., \& Wright, E.L.: First-Year Wilkinson Microwave
Anisotropy Probe (WMAP) Observations: Determination of Cosmological
Parameters, ApJS, \textbf{148}, 175--194 (2000)
\bibitem{spe07}
Spergel, D.N., Bean, R., Dore', O., Nolta, M.R., Bennett, C.L.,
Dunkley, J., Hinshaw, G., Jarosik, N., Komatsu, E., Page, L., Peiris,
H.V., Verde, L., Halpern, M., Hill, R.S., Kogut, A., Limon, M.,
Meyer, S.S., Odegard, N., Tucker, G.S., Weiland, J.L., Wollack, E.,
\& Wright, E.L.: Three-Year Wilkinson Microwave Anisotropy Probe (WMAP)
Observations: Implications for Cosmology, ApJS, \textbf{170}, 377--408 
(2007)
\bibitem{twv90} 
Tyson, J.A., Wenk, R.A., \& Valdes, F.: Detection of
systematic gravitational lens galaxy image alignments - Mapping dark
matter in galaxy clusters, ApJL, \textbf{349} 1--4 (1990)
\bibitem{vik98}
Vikhlinin, A., McNamara, B.R., Forman, W., Jones, C., Quintana, H.,
\&  Hornstrup, A.: A Catalog of 200 Galaxy Clusters Serendipitously 
Detected in the ROSAT PSPC Pointed Observations, ApJ, \textbf{502},  
558--581 (1998)
\bibitem{vik03}
Vikhlinin, A., Voevodkin, A., Mullis, C.R., VanSpeybroeck, L., 
Quintana, H., McNamara, B.R., Gioia, I.M.,  Hornstrup, A., Henry, J.P.,
Forman, W.R., \&  Jones, C.:  Cosmological Constraints from the Evolution 
of the Cluster Baryon Mass Function at z$\sim$0.5, ApJ,  \textbf{590}, 
15--25 (2003)
\bibitem{vik06}
Vikhlinin, A., Kravtsov, A., Forman, W., Jones, C., Markevitch, M., 
Murray, S.S., \& Van Speybroeck, L.: Chandra Sample of Nearby Relaxed 
Galaxy Clusters: Mass, Gas Fraction, and Mass-Temperature Relation, ApJ, 
\textbf{640}, 691--709 (2006)
\bibitem{voe04}
Voevodkin, A. \& Vikhlinin, A.: Constraining Amplitude and Slope of 
the Mass Fluctuation Spectrum Using a Cluster Baryon Mass Function, ApJ, 
\textbf{601}, 610--620 (2004)
\bibitem{voit05a}
Voit, G. M.: Tracing comic evolution with clusters of galaxies,
Rev. Mod. Phys., \textbf{77}, 207-258, (2005a)
\bibitem{voit05b}
Voit, G. M.:  Expectations for evolution of cluster scaling relations, AdSpR,
\textbf{36}, 701--705 (2005b)
\bibitem{zh61}
Zwicky, F. et al.: Catalogue of Galaxies and of Clusters 
of Galaxies, Calif. Inst. of Techn., Pasadena, 6 vols. (1961-68)  
%
\end{thebibliography}
